\documentclass[twocolumn,prb,showpacs]{revtex4}
\usepackage{amsmath}
\usepackage{color}
\usepackage{graphicx}
\usepackage{epsfig}

\makeatother
\begin{document}

\title{Transfer-matrix renormalization group study of the spin ladders \\with cyclic
four-spin interactions}
\author{H. T. Lu,$^{1,2}$ L. Q. Sun,$^{2}$ Shaojin Qin,$^{2}$ Y. J. Wang$^{3}$}

\affiliation{${}^{1}$School of Physics, Peking University, Beijing 100871, China\\
${}^{2}$Institute of Theoretical Physics and Interdisciplinary Center
of Theoretical Studies, Chinese Academy of Sciences, Beijing 100080, China \\
${}^{3}$Department of Physics, Beijing Normal University, Beijing 100875, China}

\begin{abstract}
The temperature dependence of the specific heat and spin
susceptibility of the spin ladders with cyclic four-spin
interactions in the rung-singlet phase is explored by making use
of the transfer-matrix renormalization group method. The values of
spin gap are extracted from the specific heat and susceptibility,
respectively. It is found that for different relative strength
between interchain and intrachain interactions, the spin gap is
approximately linear with the cyclic four-spin interaction in the
region far away from the critical point. Furthermore, we show that
the dispersion for the one-triplet magnon branch can be obtained
by numerically fitting on the partition function.
\end{abstract}


\maketitle

\section{Introduction\label{sec:Introduction}}

In the past few years, cyclic spin exchanges in quantum
antiferromagnetic magnets have attracted considerable attention. From
various experiments, it has been realized that four-spin cyclic
exchanges play an important role in understanding the spin dynamics of
insulating cuprate materials, such as the two-leg spin ladders
Cu$_2$O$_3$ in the compounds SrCu$_{2}$O$_{3}$,~\cite{SrCu2O3} (Sr, Ca,
La)$_{14}$Cu$_{24}$O$_{41}$,~\cite{LaCaCuO_1,LaCaCuO_2,LaCaCuO_3} and
the CuO$_{2}$ plane in La$_{2}$CuO$_{4}$.~\cite{LaCuO_1,LaCuO_2} Similar
mechanism of multiple spin exchanges in $^{3}$He (Ref.\onlinecite{He3}) and Wigner
crystal,~\cite{Wigner} which was found to be large, has also been
proposed.

Spin ladders can be viewed as intermediates between one-dimensional
spin chains and two-dimensional spin lattices. They are particular
of interest in the understanding of the high-temperature
superconductivity in two dimensions and the Haldane's
conjecture~\cite{Haldane2,Roux} in spin chains. The two-leg spin
ladder is the simplest system on which cyclic exchanges can be
materialized. In the absence of cyclic exchanges, with nonzero
interchain couplings, the $S=1/2$ spin ladder lies in a Haldane
phase, which has a short-range resonating-valence-bond (RVB) ground
state with a spin gap, and superconductivity emerges upon
doping.\cite{Ladder_1, Ladder_12} Contrary to the case in a $S=1/2$
chain, the spinons in the ladder are confined back to the
well-defined quasiparticles (massive magnons) in the neighborhood of
$q=\pi$, where a spectral gap exists.\cite{Ladder_2} It has been
shown that four-spin cyclic exchanges can make the spin gap
decreases rapidly.\cite{LaCaCuO_1} Furthermore, tuning of the
four-spin interactions can drive the system into different
phases.~\cite{Lauchli,Gritsev} At the nearest critical point, the
system becomes gapless to cross from the RVB rung-singlet phase to a
staggered dimer phase. By a field theory using the Majorana Fermion
representation, Nersesyan and Tsvelik\cite{Nersesyan&Tsvelik}
referred to the critical phenomenon as the level $k=2$ SU(2)
Wess-Zumino-Novikov-Witten model or SU(2) $c=\frac{3}{2}$ conformal
field theory. In other words, the phase transition is of the
second-order type and belongs to the universality class of the
Takhtajan-Babujian critical point.\cite{TakhtajanBabujian} This
statement has been confirmed by numerical
simulations.~\cite{Hijii_1,Hijii_2}

\begin{figure}
\centering
\includegraphics[width=7.5cm]{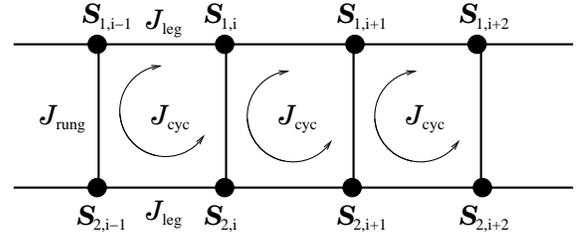}
\caption{\label{ladder}Schematic structure of a two-leg ladder with cyclic
(four-spin) interaction. The round arrows stand for the action of
the four-spin terms as given in Eq. (\ref{eq:1}), which is not completely
identical to the usual cyclic permutations.}
\vspace{0.3cm}
\end{figure}

The Hamiltonian describing an isotropic $S=1/2$ spin ladder with
additional cyclic (four-spin) interaction is represented  by
\begin{align}
H & =\sum_{i}\left\{ J_{\text{rung}}\mathbf{S}_{1,i}\cdot\mathbf{S}_{2,i}+J_{\text{leg}}\left[\mathbf{S}_{1,i}\cdot\mathbf{S}_{1,i+1}+\mathbf{S}_{2,i}\cdot\mathbf{S}_{2,i+1}\right]\right.\nonumber \\
 & +2J_{\text{cyc}}\left[\left(\mathbf{S}_{1,i}\cdot\mathbf{S}_{1,i+1}\right)\left(\mathbf{S}_{2,i}\cdot\mathbf{S}_{2,i+1}\right)\right.\nonumber \\
 & +\left(\mathbf{S}_{1,i}\cdot\mathbf{S}_{2,i}\right)\left(\mathbf{S}_{1,i+1}\cdot\mathbf{S}_{2,i+1}\right)\nonumber \\
 & \left.\left.-\left(\mathbf{S}_{1,i}\cdot\mathbf{S}_{2,i+1}\right)\left(\mathbf{S}_{2,i}\cdot\mathbf{S}_{1,i+1}\right)\right]\right\} , \label{eq:1}
\end{align}
where the indices 1 and 2 distinguish upper and lower legs, and
$i$ labels rungs; $J_{\text{rung}}$ and $J_{\text{leg}}$ are the
interchain and intrachain couplings, respectively;
$J_{\text{cyc}}$ denotes the strength of the cyclic (four-spin)
exchange. The model is schematically shown in Fig. \ref{ladder}.
In the rest of the paper, for convenience, we call Eq.
(\ref{eq:1}) as type-I Hamiltonian.

The above Hamiltonian distinguishes from another one where the
ring exchange term is directly represented by a cyclic permutation
operator $P_{ijkl}$ in the plaquette:
\[
K\left(P_{ijkl}+P_{ijkl}^{-1}-\frac{1}{4}\right). \label{eq:P}
\]
We call this form as type-II Hamiltonian. The cyclic permutation
operator can be expressed by spin-$\frac{1}{2}$ operators
including bilinear frustrating and biquadratic
terms.~\cite{LaCaCuO_1,Mikeska2} So type-II Hamiltonian can be
re-written as
\begin{align}
H & =\sum_{i}\left\{ \left(J_{\text{rung}}+2K\right)\mathbf{S}_{1,i}\cdot\mathbf{S}_{2,i}\right.\nonumber \\
 & +\left(J_{\text{leg}}+K\right)\left(\mathbf{S}_{1,i}\cdot\mathbf{S}_{1,i+1}+\mathbf{S}_{2,i}\cdot\mathbf{S}_{2,i+1}\right)\nonumber \\
 & +K\left(\mathbf{S}_{1,i}\cdot\mathbf{S}_{2,i+1}+\mathbf{S}_{2,i}\cdot\mathbf{S}_{1,i+1}\right)\nonumber \\
 & +4K\left[\left(\mathbf{S}_{1,i}\cdot\mathbf{S}_{1,i+1}\right)\left(\mathbf{S}_{2,i}\cdot\mathbf{S}_{2,i+1}\right)\right.\nonumber \\
 & +\left(\mathbf{S}_{1,i}\cdot\mathbf{S}_{2,i}\right)\left(\mathbf{S}_{1,i+1}\cdot\mathbf{S}_{2,i+1}\right)\nonumber \\
 &
 \left.\left.-\left(\mathbf{S}_{1,i}\cdot\mathbf{S}_{2,i+1}\right)\left(\mathbf{S}_{2,i}\cdot\mathbf{S}_{1,i+1}\right)\right]\right\}.
\label{eq:2}
\end{align}

There have been extensive theoretical studies on the type-II
Hamiltonian.  By the density matrix renormalization group (DMRG)~\cite{White}
and exact diagonalization method, the energy spectra about one-magnon
branch and two-magnon continuum in the rung-singlet phase have been
obtained.~\cite{LaCaCuO_3, Roux} The phase diagram for
$J_{\text{rung}}=J_{\text{leg}}$ was studied in detailed~\cite{Lauchli} and a
global picture for general cases has been proposed.~\cite{Gritsev} The phase
transitions, mainly concerning the case from the rung-singlet to the
dimerized phase, together with the critical exponent were studied by field theory,
perturbative and series expansion method.~\cite{Hijii_1, Hijii_2, Mikeska2,
Schmidt} 

On the other hand, the type-I Hamiltonian catches less attention. And it is
widely believed that the physics should be the same for the type-I and type-II
Hamiltonians, although there is a lack of detailed comparisions. Theoretically, the ring
exchange interaction naturally arises as an effective low-energy model for the
multiband Hubbard model near the insulating filling and in the strong coupling
limit. It has been shown that the four-spin terms are most significant
ones.\cite{Muller} In order to highlight the effect of this four-spin term, in
this paper, we employ the type-I Hamiltonian for our numerical simulations.

As mentioned above, the zero-temperature behaviors of the spin ladders with
cyclic exchange interactions have been analyzed in detail, especially for the
type-II Hamiltonian. A study on the thermodynamic properties of the systems
has not been carried out extensively, except for the study by a
high-temperature series expansion method.~\cite{Thermo_1} These quantities can
be directly compared with the experimental observations and therefore are of
particular use. Detailed and reliable numerical results on these quantities
may offer an expedient way to determine the model parameters for the compounds
we are interested in. In this paper, we show that by using the transfer-matrix
renormalization group (TMRG) method, we can not only obtain the temperature
dependence of the specific heat and susceptibility of the spin ladders down to
the low-temperature regime and with high accuracy, but also extract extra
information from the data for the magnitude of the gap and the one-triplet
magnon excitations, which can be compared with the experiments directly.

As proposed by experiments, for the insulator cuprate ladder
materials, ~\cite{Windt01,SrCu2O3,LaCaCuO_1,LaCaCuO_2,LaCaCuO_3}
both $J_{\text{rung}}$ and $J_{\text{leg}}$ are positive and close
in magnitude. The values of $K/J_{\text{rung}}$ range in a region
$0.025\le K/J_{\text{rung}}\le0.075$, which roughly corresponds to
the parameter region $0.05\le J_{\text{cyc}}/J_{\text{rung}}\le
0.15$ for the type-I Hamiltonian and the system is still in the
rung-singlet phase. For this reason, in our numerical simulations,
we confine the parameter $x_{\text{cyc}}\in[0,0.1]$, and $x$ is
taken three different values: 0.5, 1.0 and 1.2, where
\begin{eqnarray}
x & := & J_{\text{leg}}/J_{\text{rung}},\label{eq:x} \\
x_{\text{cyc}} & := & J_{\text{cyc}}/J_{\text{rung}}.
\label{eq:xcyc}
\end{eqnarray}
We take $x=0.5$ in order to make a comparison with the typical
rung-singlet phase.

The numerical algorithm we employ to study the thermodynamical properties of
the present model is the TMRG
method,~\cite{Bursill2,XiangWang2,XiangWang} which is a powerful numerical
tool for studying the thermodynamic properties of one-dimensional quantum
systems. One is able to evaluate nearly all thermodynamic quantities by the
maximum eigenvalue and the corresponding left and right eigenvectors of the
transfer matrix. We will show later that the magnitude of the gap and the
low-lying excitations can also be extracted from the obtained data. In our
numerical calculations, $\tau =0.05$, the error caused by the Trotter-Suzuki
decomposition is less than $10^{-3}$. During the TMRG iterations, $80-100$
states are retained and the truncation error is less than $10^{-4}$ down to
$k_{B}T\sim0.01J_{\text{rung}}$.

\section{Specific heat and spin susceptibility\label{sec:Numerical}}

\begin{figure}
\includegraphics[width=7.5cm]{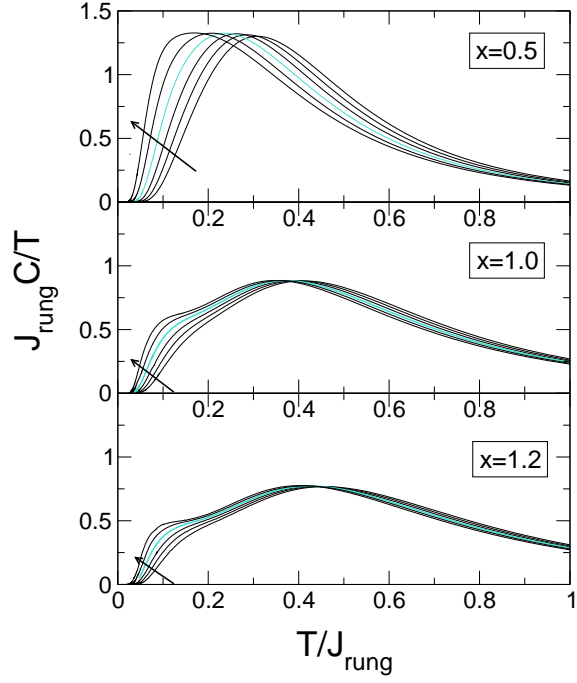}
\centering
\caption{\label{C/T}(Color online) The specific heat coefficient $C/T$ for
type-I Hamiltonian
for three values of $x$ with $x_{\text{cyc}}=$ 0,
0.02, 0.04, 0.06, 0.08, and 0.1, in ascending order along the direction
of arrows.}
\vspace{0.3cm}
\end{figure}

Figure \ref{C/T} shows the TMRG results on the temperature
dependence of the specific heat coefficient $C/T$ for $x=$ 0.5, 1.0,
1.2 with $x_{\text{cyc}}=$ 0, 0.02, 0.04, 0.06, 0.08 and 0.1.
Besides the exponential decrease as $T\to 0$, which indicates the
existence of a gap, the general behavior of the curves for a given
$x$ is as follows: As $x_{\text{cyc}}$ increased, the slope of
$(C/T)_{\text{max}}$ shifts to lower temperatures. The reason is
obvious: that with the increase of the four-spin cyclic exchange, the
spin gap and the whole dispersion decrease.~\cite{LaCaCuO_1,Schmidt}
Another interesting feature is that at $x=$ 1.0 and 1.2, a shoulder
gets more and more distinct at low temperatures with the increase of
$x_{\text{cyc}}$. We attribute this shoulder to the tendency of
deconfinement of bound spinons from the magnons, as the crossover
occurs from the strong coupling regime
$\left(J_{\text{leg}}/J_{\text{rung}}=x\ll1\right)$ to the weak
coupling one $\left(J_{\text{leg}}/J_{\text{rung}}=x\gg1\right)$ with the
extreme limiting case of a finite $C/T$ value at $T=0$.

\begin{figure}
\includegraphics[width=7.5cm]{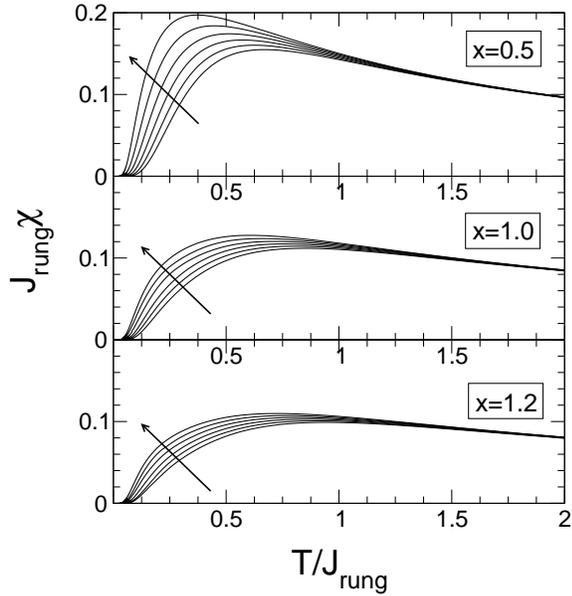}
\centering
\caption{\label{chi_1}Temperature dependence of the susceptibility $\chi$
for type-I Hamiltonian for three values of $x$ with $x_{\text{cyc}}=$
0, 0.02, 0.04, 0.06, 0.08, and 0.1 in ascending order along the direction
of arrows.}
\vspace{0.5cm}
\end{figure}

The corresponding results for spin susceptibility $\chi(T)$ are
presented in Fig.~\ref{chi_1}. The exponential decrease of $\chi(T)$
as $T\rightarrow0$, and the shift of the peak to lower temperatures
with the increase of $x_{\text{cyc}}$ are consistent with the above
behaviors of $C/T$. Our numerical result on the susceptibility well
agrees with that of Ref. \onlinecite{Thermo_1} obtained by
high-temperature series expansion and exact diagonalization.

\begin{figure}
\includegraphics[width=7.5cm]{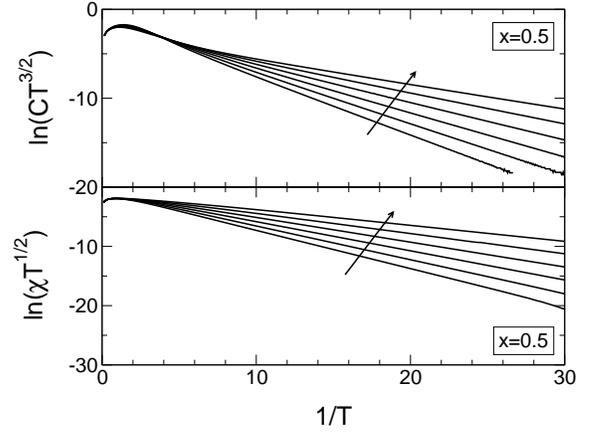}
\centering
\caption{\label{0.5g}$1/T$ dependence of $\ln\left(CT^{3/2}\right)$ and
$\ln\left(\chi T^{1/2}\right)$ (in the unite of $J_{\text{rung}}$)
for $x=0.5$ with $x_{\text{cyc}}=$ 0, 0.02, 0.04, 0.06, 0.08, and
0.1 in ascending order along the direction of arrows.}
\vspace{0.3cm}
\end{figure}

The $x_{\text{cyc}}$ dependence of the gap is a central quantity,
since basically it determines the relevancy of the cyclic exchange
term. In the vicinity of the critical point $x_{\text{cyc}}^c$,
generally,
\[\Delta\sim\mid x_{\text{cyc}}-x_{\text{cyc}}^c\mid ^\eta. \label{eq:eta}\]
For the type-II Hamiltonian in both strong and weak coupling limits,
by the perturbative cluster expansion and Pad\'e approximation in
the strong coupling regime, together with the bosonization technique
in the weak coupling limit, M\"uller \emph{et al.}\cite{Mikeska2} showed
that the critical exponent $\eta$ for the gap is approximately equal
to 1; i.e., the gap is linearly dependent on $x_{\text{cyc}}$ near
the critical point. For the $J_{\text{leg}}=J_{\text{rung}}$ case,
this result was also obtained by Hijii \emph{et al.}\cite{Hijii_2} Here, we
will see that for type-I Hamiltonian, in the region considerably far
away from the critical point, $x_{\text{cyc}}\in(0,0.1)$, the linear
behavior of the gap vs. $x_{\text{cyc}}$ still holds in almost all
the parameter range.

\begin{figure}
\includegraphics[width=7.5cm]{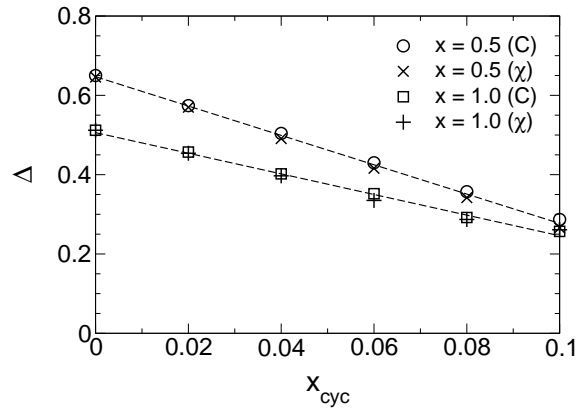}
\centering \caption{\label{gap}The $x_{\text{cyc}}$ dependence of
the spin gap for $x=$ 0.5 and 1.0 obtained by fitting the Eqs.
(\ref{eq:4}) (signed as $C$) and (\ref{eq:5}) (given as $\chi$),
respectively.} \vspace{0.6cm}
\end{figure}

For the spin-$\frac{1}{2}$ ladder in the rung-singlet phase, similar
to the case in a $S=1$ chain, we can assume that the low-lying
excitation spectrum 
$\epsilon(k)$ for $k$ near $\pi$ is approximately given by 
\begin{eqnarray}
\epsilon(k)&=&\sqrt{\upsilon^2(k-\pi)^2+\Delta^2}\nonumber\\
&\approx&\Delta+\frac{\upsilon^2}{2\Delta}(k-\pi)^2+\mathcal{O}\left(|k-\pi|^3\right),
\label{eq:quadraticdispersion}
\end{eqnarray}
where $\upsilon$ is the spin wave velocity and $\Delta$ is the spectrum gap.
The low temperature behaviors of the
specific heat and susceptibility per spin can be expressed, respectively,
as~\cite{Xiang}

\begin{align}
C\left(T\right) & \approx
\frac{3\Delta}{\upsilon\sqrt{8\pi}}\left(\frac{\Delta}{T}\right)^{3/2}e^{-\Delta/T}
, \label{eq:4}\\
\chi\left(T\right) & \approx \frac{1}{\upsilon}\sqrt{\frac{\Delta}{2\pi
T}}e^{-\Delta/T} , \label{eq:5}
\end{align}
for $T\ll\Delta$. In this situation,
$\ln\left(CT^{3/2}\right)$ and $\ln\left(\chi T^{1/2}\right)$ should
be linear to $1/T$ with a slope $-\Delta$. Figure \ref{0.5g} shows
the $1/T$ dependence of $\ln\left(CT^{3/2}\right)$ and
$\ln\left(\chi T^{1/2}\right)$ (in units of $J_{\text{rung}}$) for
$x=0.5$ with $x_{\text{cyc}}=$ 0, 0.02, 0.04, 0.06, 0.08, and 0.1.
We observe a well-behaved linearity at $1/T>10$. For $x=$ 1.0 and
1.2, we have the similar results. The values of the gap at various
$x$ and $x_{\text{cyc}}$ can thus be extracted by fitting the data
in the interval $\beta=1/T\in(10, 20)$. The results for $x=0.5$ and
$x=1.0$ are presented in Fig. \ref{gap}. The gaps obtained from the
specific heat and susceptibility are well coincident with each
other.

\begin{figure}
\includegraphics[width=7.5cm]{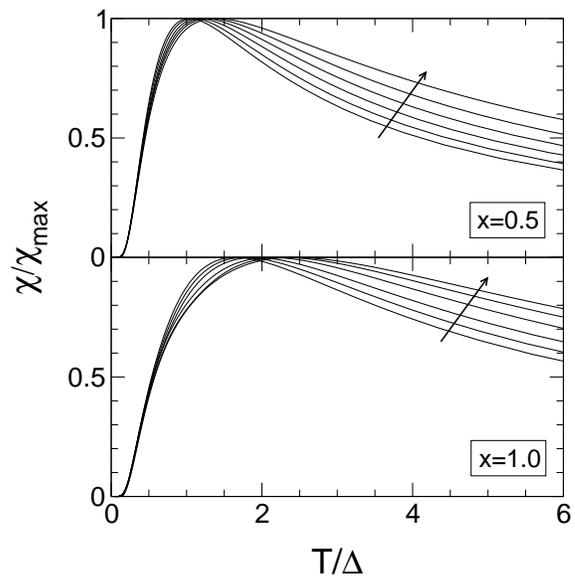}
\centering
\caption{\label{chi_2}$\chi/\chi_{\text{max}}$ versus $T/\Delta$ for
$x=0.5$, $1.0$ with $x_{\text{cyc}}=$ 0, 0.02, 0.04, 0.06, 0.08,
and 0.1 in ascending order along the direction of arrows.}
\vspace{0.3cm}
\end{figure}

To both verify the low-temperature scaling behavior of the system and check
the validity of Eq. (\ref{eq:5}), in Fig.~\ref{chi_2}, the rescaled
susceptibility $\chi/\chi_{\text{max}}$ versus $T/\Delta$ for $x=0.5$, $1.0$
at various $x_{\text{cyc}}$ is presented. For a fixed value of $x$, the curves
converge to a single one at low temperatures, demonstrating the dominance of
$\Delta$ in determining the susceptibility at low temperatures. Furthermore,
it also suggests that the values of the gap we obtained by this approach are
quantitatively reliable.  This situation is quite similar to the case of the
$S=1$ antiferromagnetic Heisenberg chain with single-ion
anisotropy.~\cite{Coombes98} Here $\chi_{max}$ serves as a common factor which
incorporates the unknown effect on the prefactor in Eq.~(\ref{eq:5}).

From Fig. \ref{gap}, we observe that the gap is approximately
linearly dependent on $x_{\text{cyc}}$ for fixed $x$ in the
parameter space we studied. We also notice the different descending
rates of $\Delta$ with $x_{\text{cyc}}$ for various $x$. At
$x_{\text{cyc}}=0$, $\Delta(x=0.5)>\Delta(x=1.0)$ with the
difference $\sim 0.15$. This indicates that the gap decreases as
the intrachain coupling $J_{\text{leg}}$ increases, which is
consistent with the perturbative picture from the dimer limit of
uncoupled rungs (strong coupling regime). On the other hand, at
$x_{\text{cyc}}=0.1$, the difference of the gaps becomes very small.
We expect for higher value of $x_{\text{cyc}}$ a reverse of the gaps
with $\Delta(x=0.5)<\Delta(x=1.0)$ should occur.

For $x=1.0$ and $1.2$ (not shown in Fig.~\ref{gap}), the values of
the gap obtained by the present method well coincide with the
numerical results from the density-matrix renormalization group
(DMRG) study for the type-II Hamiltonian for the corresponding
values of the cyclic exchange coupling.\cite{Ladder_1, LaCaCuO_3}
For instance, when $x=1.0$ and $x_{\text{cyc}}=0$, the gap obtained
here by TMRG is $\sim 0.51$, as compared with the DMRG result of
$0.504$.\cite{Ladder_1}

To come to a close, we estimates the errors in the process of gap fitting. On
the one hand, as mentioned in the previous section, combining the errors from
the Trotter-Suzuki decomposition with the TMRG truncation, we estimate that
the error caused by the TMRG algorithm in the fitting interval (in our work
$T\in(0.05, 0.1)$) is less than $10^{-2}$ at most. On the other hand, in
obtaining the fitting formulas (\ref{eq:4}) and (\ref{eq:5}), the
low-temperature limit and quadratic approximation (See
Eq.~(\ref{eq:quadraticdispersion})) have been adopted. For the specific heat,
the expression including the higher order contributions of $T/\Delta$ is given
explicitly as~\cite{Troyer}
\begin{align}
C(T)=&\frac{3\Delta}{\upsilon\sqrt{8\pi}}
\left(\frac{\Delta}{T}\right)^{3/2} \nonumber\\
&\times\left[1+\frac{T}{\Delta}+\frac{3}{4}\left(\frac{T}{\Delta}\right)^{2}\right]
e^{-\Delta/T}.
\end{align}
The uncertainty in Eq. (\ref{eq:4}) is roughly estimated as 
\begin{equation}
\delta\Delta \sim
\bar{T}\ln\left[1+\frac{\bar{T}}{\Delta}+\frac{3}{4}\left(\frac{\bar{T}}{\Delta}\right)^2\right],
\label{eq:delta}
\end{equation}
where $\bar{T}$ represents some value of temperature $T$ lying in the fitting
interval. From this, in our case the uncertainty of gap due
to the fitting formulas can be estimated approximately as $\delta\Delta\approx
0.01-0.02$. Based on these considerations, we conclude that the error in the gap
fitting is about $10^{-2}$, which is not exceeding the symbol size in Fig.
\ref{gap}.

\section{Partition function and the dispersion relations \label{sec:Numerical2}}

In TMRG calculations, the partition function
$Z_{\text{TMRG}}(\beta)$ is the original and the most accurate
numerical quantity. In previous section we obtained the
specific heat and spin susceptibility, which can also be got from
the derivatives of the partition function directly. In the following
we will discuss the single-particle state in this system and show
that its dispersion, as well as the ground-state energy and the
magnitude of the gap, can be obtained from the analyses of the
partition function.

In the rung singlet phase, the elementary excitations are gapped.
The energy gap and dispersion for the single particle state can be
analyzed perturbatively.~\cite{Mikeska2} When $J_{\text{leg}} =
J_{\text{cyc}} = 0$, the system is consisted of an array of rungs
with a vacuum state being the product of spin singlet state on each
rung.  Above this vacuum state, a hard core boson of spin 1 is the
only possible excited state on each rung. The system has a flatband
dispersion $\epsilon(k)=J_{\text{rung}}$ for single particle
excitations.  Due to the hard core scattering between the bosons,
the statistics of the system is essentially Fermi-Dirac
type,~\cite{Jizhong} which allows us to write down the partition
function for the system:
\begin{align}
\ln Z_{\text{eff}}(\beta) & = & \frac{1}{2\pi}\int_{-\pi}^{\pi}
\ln\left( 1+3 e^{-\beta\epsilon(k)} \right) dk-\beta e_0, \label{eq:zeff}
\end{align}
where $e_0$ is the ground-state energy per rung in the thermodynamic
limit and the prefactor $3$ before $e^{-\beta \epsilon(k)}$
comes from the three internal states of the single particle
excitations.

In the rung singlet phase, switching upon the couplings
$J_{\text{leg}}$ and $J_{\text{cyc}}$ will induce a
dispersion~\cite{Mikeska2,Roux,LaCaCuO_3} for the single particle
state, and the interaction between the particles. Classified by the
internal symmetries, the two-particle interaction is either
spin-channel independent or spin-channel dependent. A bound state may
also appear in certain spin-channels. This has been observed in a
previous study,~\cite{LaCaCuO_3} in which it is shown that the bound
state is also momentum dependent. Therefore, the interaction is very
complicated and depends on the total momentum of the scattering
particles.

\begin{figure}
\centering
\includegraphics[width=7.5cm]{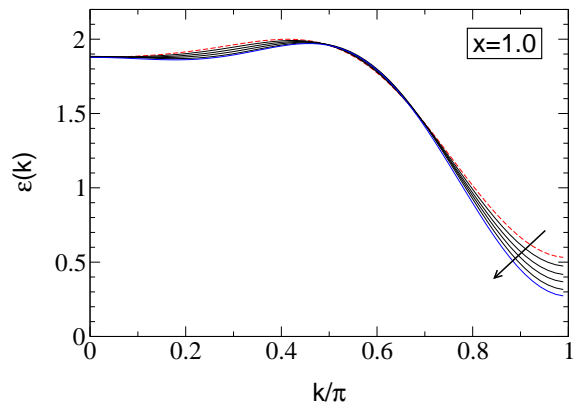}
\caption{\label{disper}(Color online) Dispersion (in units of $J_{\text{rung}}$)
for $x=1.0$ with $x_{\text{cyc}}=$ 0, 0.02, 0.04,
0.06, 0.08, and 0.1 in descending order along the direction of arrows.}
\vspace{0.3cm}
\end{figure}

Since we are dealing with a system with massive particles at zero
chemical potential, the density of the particles would be dilute at
low temperatures. Taking into account of the form of the effective
partition function for single particle dispersion $Z_{\text{eff}}$,
we expect it would match the accurate TMRG results at both high and
low temperatures by including higher order harmonics in the
dispersion:~\cite{Troyer}
\begin{align}
\epsilon(k) & = &
\mu_0 + \mu_1 \cos(k) + \mu_2 \cos(2k) + \mu_3 \cos(3k)
. \label{eq:eps1}
\end{align}
The energy gap is given by $\Delta = \mu_0 - \mu_1 + \mu_2 - \mu_3$
at $k=\pi$. Starting with a set of trial values for $\mu$'s given by
perturbative calculation,~\cite{Mikeska2} we minimize the squared
2-norm of the residual in free energy calculated from the trial
dispersion:
\begin{align}
W & (\mu_0,\mu_1,\mu_2,\mu_3,e_0)  =   \nonumber \\
& \sum_{i=1}^N\left[ \ln Z_{\text{eff}}\left(\beta_i\right) - \ln
Z_{\text{TMRG}} \left( \beta_i\right)\right]^2,
\label{eq:error}
\end{align}
where $\beta_i$ are smoothly distributed in the interval $(5,15)$
with a spacing $\Delta \beta=0.05$. Typically, we show the fitting
results for the $x=1.0$ case in Table \ref{Tab_1.0}. The
corresponding dispersions are plotted in Fig. \ref{disper}. In our
final numerical results, the values of $W$ in Eq. (\ref {eq:error})
are of order $10^{-6}$, and the mean deviation at each data point is
less than $10^{-4}$.

\begin{table}
\centering

\tabcolsep 0.22cm
\begin{tabular}{cccccccc}
\hline
\hline
$x_{\text{cyc}}$& $\mu_{0}$& $\mu_{1}$& $\mu_{2}$& $\mu_{3}$& $e_{0}$&
$\Delta$& $\Delta_{\text{C}}$\tabularnewline
\hline
0& 1.58& 0.59& -0.38& 0.09& -1.156& 0.53& 0.51
\tabularnewline
\hline
0.02& 1.57& 0.60& -0.39& 0.10& -1.137& 0.47& 0.46
\tabularnewline
\hline
0.04& 1.55& 0.62& -0.41& 0.12& -1.119& 0.42& 0.40
\tabularnewline
\hline
0.06& 1.54& 0.63& -0.42& 0.13& -1.101& 0.37& 0.35
\tabularnewline
\hline
0.08& 1.53& 0.64& -0.43& 0.15& -1.084& 0.31& 0.30
\tabularnewline
\hline
0.1& 1.52& 0.65& -0.44& 0.15& -1.068& 0.27& 0.26
\tabularnewline
\hline
\hline
\end{tabular}

\caption{\label{Tab_1.0}The partition function fitting results
by Eq. (\ref{eq:zeff})
for the $x=1.0$ case. $\Delta_{\text{C}}$
in the last column is the energy gap obtained by Eq. (\ref{eq:4}), as plotted in Fig. \ref{gap}.}
\vspace{0.7cm}
\end{table}

From Table \ref{Tab_1.0}, we find that $\mu_0$ and $\mu_1$
monotonically decreases and increases with $x_{\text{cyc}}$,
respectively. This is consistent with the perturbative
results.~\cite{Mikeska2} On the other hand, the third order
coefficient $\mu_3$ increases with $x_{\text{cyc}}$. At
$x_{\text{cyc}}=0.1$, it reaches 0.15, about one-third of $\mu_2$.
This indicates that the approximation up to the third order in the
dispersion is necessary for $x=1.0$ case. On the contrary, for
$x=0.5$, $\mu_0 \sim 1$, $\mu_1 \sim 0.5 - 0.6$, the absolute value
of both $\mu_2$ and $\mu_3$ are less than $0.1$, and the dispersion
can be satisfactorily approximated by single-component harmonics
with only $\mu_1$. This fact reflects that with the increase of
$J_{\text{leg}}$ and $J_{\text{cyc}}$, higher orders in the harmonic
expansion are needed to describe the one-particle excitation in the
spectrum.

The dispersion relations shown in Fig. \ref{disper} should
correspond to the one-triplet magnon branch in the spectrum of the
ladder. We compare the results with the previous studies by the
DMRG.~\cite{LaCaCuO_3, Roux} We find they have quite similar shapes
and are quantitatively agreement in the vicinity of $k=\pi$. The
deviation when approaching $k=0$ can be attributed partially to the
two-triplet bound state arising from the magnon-magnon interaction
that we have ignored.~\cite{LaCaCuO_3}

The energy gaps obtained in this way are presented in Table
\ref{Tab_1.0}, and are compared with their counterparts from the
specific heat data. We see that they are rather coincident. In
addition, we have obtained the ground state energy ($e_0$ in
Eq.~(\ref{eq:zeff})), which is compatible with the results from the
zero-temperature DMRG scheme.

\section{Summary\label{sec:Summary}}

In this paper, we explored the thermodynamic properties of the spin
ladders with cyclic four-spin interactions by the TMRG method.
The temperature dependences of the specific heat and the susceptibility
of the ladder in the rung-singlet phase are analyzed numerically.
Based on the above results, we extracted the values of the spin gap
and found that it manifests approximately linear behavior with the
cyclic interaction in the parameter space we considered. We also
showed that the dispersion for low-lying excitations can be obtained
from the partition function very effectively by combining with the
mean-field or perturbative studies. Comparing with the high
temperature series expansion and the exact diagonalization
approaches,~\cite{Thermo_1} we believe that the results presented in
this study give more direct and accurate physical quantities in the
thermodynamic limit. Further experimental results on specific heat
and susceptibility may allow one to determine the various
parameters for the two-leg spin ladder materials via this approach.
In addition, the dispersion relations extracted from the free energy
may be used to be compared with the experimental data of the
inelastic neutron scattering spectra.

\begin{acknowledgments}
This work was supported by the National Natural Science Foundation of China.
The numerical work of this project was performed on the HP-SC45 Sigma-X
parallel computer of ITP and ICTS, CAS.
\end{acknowledgments}

\end{document}